\documentclass{acm_proc_article-sp}
\usepackage{amssymb}
\usepackage{amsfonts}
\usepackage{latexsym}
\usepackage{amsmath}
\usepackage{graphicx}
\usepackage{subfigure}
\usepackage{algorithm}
\usepackage{algorithmic}

\begin{document}

\title{An Adaptive Online Ad Auction Scoring Algorithm for Revenue
Maximization}
\author{Chenyang Li, Mingyi Hong, Randy Cogill and Alfredo Garcia \\
\affaddr{Department of Systems and Information Engineering}\\
\affaddr{University of Virginia, Charlottesville, Virginia}\\
\email{\{cl2ha, mh4tk, rcogill, ag7s\}@virginia.edu} } \maketitle



%
%
\textbf{Keywords:} ad auction, sponsored search, adaptive algorithms, search
engines, revenue maximization

\section{Introduction}

Sponsored search becomes an easy platform to match potential consumers'
intent with merchants' advertising. Advertisers express their willingness to
pay for each keyword in terms of bids to the search engine. When a user's
query matches the keyword, the search engine evaluates the bids and
allocates slots to the advertisers that are displayed along side the unpaid
algorithmic search results. The advertiser only pays the search engine when
its ad is clicked by the user and the price-per-click is determined by the
bids of other competing advertisers.

It seems natural to assume that the number of clicks an ad will get when
displayed depends on some ad-specific factors. For example, when a potential
buyer searches the key word \textquotedblleft tennis racquet", an ad with
\textquotedblleft The Lowest Prices Guaranteed!" seems to be more intriguing
and more likely to result in a click than an ad with \textquotedblleft Save
on Tennis Racquets". Besides, an online tennis store that has built a high
reputation among the internet shoppers is more likely to attract more clicks
than an online tennis store that keeps leaving customers negative
experiences. There are many other measures of \textquotedblleft ad quality"
search engines may also consider in deciding which ads to display and at
which positions.

These measures are incorporated into an \textquotedblleft ad quality"
scoring system that is widely used by the major search engines. For example,
Google ranks ads by bid times the corresponding ad-quality score. When the
ad-quality score of each advertiser is interpreted as the expected number of
clicks the corresponding ad can generate once it is displayed, then the ads
are ranked in the order of their expected revenue. This interpretation of
the ad-quality score conveys a message to the advertisers: the search engine
wants the higher ranked slots, those more likely to receives clicks, to go
to the ads expected to generate higher revenues for it. It is not strange
that all search engines' goal in selling the ad slots is revenue
maximization, but how do they achieve the goal through the ad-quality
scoring remains a secrete to the advertisers. A simple example can show
that, though intuitively appealing, ranking ordered by bid times expected
click-through rate might not be the optimal ranking to generate maximum
revenue for the search engine. Then two problems become interesting: What is
a good scoring strategy for the search engine to maximize its revenue? Does
the interpretation of the ad-quality score truly represents what the
ad-quality score does in the process of revenue maximization?

In this paper, we study the scoring strategy of the search engine for the
purpose of revenue maximization. Section 2 gives the preliminaries needed
for our analysis. After specifying the ad auction model rules, notations and
the related literature, we show some ad scoring examples that motivated this
work.\ In Section 3, we focus on the complete information setting, where
advertisers' values are known. Our analysis shows that when there are more
advertisers than slots, there is a scoring strategy that induces a Nash
equilibrium formed by the set of truthful bids. Under such an equilibrium,
the search engine takes all the social surplus and\ the socially optimal
ranking of the advertisers is the optimal ranking for maximizing the search
engine's revenue. Furthermore, we show that under such a truthful bidding
Nash equilibrium, when the click-through rate for an advertiser at a certain
slot takes a product form of an ad-specific factor and a position-specific
factor, there exists a $0\left( N^{2}\right) $ time algorithm to find the
socially optimal ranking of the advertisers which also maximize the search
engine's revenue.

In Section 4, we move to the incomplete information case where advertisers'
values are not known by the search engine. Based on some rational
advertisers behavior arguments, we propose an adaptive online algorithm that
dynamically scores ads to reveal advertisers' values and to manipulate their
rankings to maximize the search engine's total revenue. In Section 5, we
test the algorithm under the 8-slot sponsored ads setting used by Google,
Microsoft (Bing.com) and Yahoo! and simulation results show that the
performance of the algorithm agrees very well with our theoretical analysis.
We conclude the paper in Section 6.

\section{Preliminaries}

\subsection{Auction rules}

We focus on a single keyword slot auction, where there are $N$ advertisers
competing for $S$ slots. Without loss of generality, we assume $S\geq N$,
since redundant slots can remain blank. Let $i=1,...,N$ index advertisers
and $j=1,...,S$ index slots. Let $v_{i},b_{i}$ be the value, bid of
advertiser $i$ for a particular keyword and $p_{i,j}$ be the price per click
advertiser $i$ pays at slot $j$. Let $e_{i}\,>0$\ be the ad-quality score
the search engine assigns to advertiser $i$ to represent a measure of the
predicted click-through rate advertiser $i$ will generate and $x_{i,j}$ be
the actual click-through rate advertiser $i$ receives at slot $j$ per hour.

We follow the rules of the auction described in \cite{Var09} by Varian that
are used by the major search engines.

(1) Each advertiser $i$ chooses a bid $b_{i}$.\

(2) The advertisers are ordered by bid times predicted click-through rate $%
b_{i}\times e_{i}$.

(3) The price that advertiser $i$ pays for a click is the minimum necessary
to retain its position.

(4) If there are fewer bidders than slots, the last bidder pays a reserve
price $p_{r}$.

\subsection{Previous literature}

When $e_{i}\equiv 1$ for all the advertisers, the above described auction
becomes the Generalized Second Price Auction (GSP) first developed and
adopted by Google in 2002, where advertisers are ranked by their bids and
the advertiser assigned slot $j$ pays the price per click equal to the bid
of the advertiser assigned slot $j+1$. Early studies on some basic
properties of GSP are independently reported by Edelman et al. \cite%
{Edelman07} and Varian \cite{Var07}.

In their analysis, the ad auction problem is modeled as a game where each
advertiser is a player who uses bidding as strategy to maximize its surplus,
i.e., the value of the clicks it receives minus the price it pays the search
engine for those clicks. The game becomes of special interest when in
equilibrium, each advertiser prefer its current slot to other slots and has
no incentive to change how it is bidding. Here is the formal definition.

{\normalsize \textbf{Definition 1}. Given the advertisers' value $\left(
v_{1},...,v_{N}\right) $, a Nash equilibrium is the set of bids so that
given these bids, no advertiser has an incentive to change its bid.
Specifically, a Nash Equilibrium (NE) satisfies, for any advertiser $i$,
\begin{equation*}
\left( v_{i}-p_{i,j}\right) x_{i,j}\geq \left( v_{i}-p_{i,j^{\prime
}}\right) x_{i,j^{\prime }}\text{ \ \ for all slots }j^{\prime }\neq j.
\end{equation*}%
}

Varian in \cite{Var07} studies the bidding behavior under the GSP auction
and shows every advertiser has a range for the bid to place to maintain its
current position. He derives many meaningful results when the advertisers'
bidding profile and the corresponding pay per click prices reach a subset of
Nash Equilibrium, namely the symmetric Nash equilibrium (SNE). Based on the
lower and upper bounds of the SNE bids, Varian derives the lower and upper
bounds of the total revenue a search engine can obtain \cite{Var07} and
further estimates advertisers' surpluses based on the prices per click they
pay \cite{Var09}.

Lahaie \cite{Lahaie06} gives a good study on the two auctions, one with all $%
e_{i}$ being equal and the other with $e_{i}$ representing the expected
click-through rate and names them \textquotedblleft rank by bid" and
\textquotedblleft rank by revenue" respectively. Both complete and
incomplete settings are considered and estimation of the search engine's
revenue is given. However, there is no analysis of revenue-maximizing
scoring strategy for the searching engine.

The most relevant work to ours can be found in Liu and Chen \cite{Liu06}
where the \textquotedblleft ad-quality" score is referred to as the
\textquotedblleft weighting factor". They study in the incomplete
information setting both the \textquotedblleft efficient" weighting factor
that maximize the total social surplus and the \textquotedblleft optimal"
weighting factor that maximize the search engine's revenue. However, they
only focus on the ad auction model with one slot and the quality type for
the advertisers is binary (high or low), while we study multiple advertisers
and slots and obtain the corresponding results through equilibrium analysis
in the complete information setting.

\subsection{Motivating Examples}

In \cite{Var07} Varian gives the lower and upper bounds of the total revenue
a search engine can obtain by selling the ad positions but does not discuss
whether this revenue under the symmetric Nash equilibrium is optimal. Here,
a simple example shows that when the ad-quality score $e_{i}$ accurately
predicts the click-through rate for each advertiser $i$, the resulting
ranking based on $b_{i}\times e_{i}$ might not generate the maximum revenue
for the search engine.

Table 1 gives the different click-through rates three advertisers Coke,
Pepsi and Dr. Pepper will get under different rankings for the key word
\textquotedblleft soda drink" in a 3-slot ad auction. Let us assume that
Coke and Pepsi are the only two advertisers that sell cola and potential
buyers looking for cola will either click on Coke or Pepsi. Further assume
Coke has a bigger brand name and whenever placed at the same position will
always receive a higher click-through rate than Pepsi ($%
x_{Coke,j}>x_{Pepsi,j}$).
\begin{table*}[t]
{\small \vspace{-0.1cm}
\begin{equation*}
\begin{tabular}{c|c|c|c|c|c|c|c|c|c|c|c|c}
\multicolumn{13}{c}{Table 1: Click-through rates under different rankings
for Coke, Pepsi and Dr. Pepper.} \\ \hline
Advertiser & Rank & $x_{i,j}$ & Rank & $x_{i,j}$ & Rank & $x_{i,j}$ & Rank &
$x_{i,j}$ & Rank & $x_{i,j}$ & Rank & $x_{i,j}$ \\ \hline\hline
Coke & 1 & 70 & 1 & 80 & 2 & 50 & 2 & 70 & 3 & 40 & 3 & 50 \\ \hline
Pepsi & 2 & 30 & 3 & 20 & 1 & 50 & 3 & 30 & 1 & 60 & 2 & 50 \\ \hline
Dr. Pepper & 3 & 20 & 2 & 30 & 3 & 20 & 1 & 35 & 2 & 30 & 1 & 35 \\ \hline
\end{tabular}%
\end{equation*}
\vspace{-0cm}}
\end{table*}

{\normalsize Tables 2-4 give the advertisers' bids, surpluses and the search
engine's revenue generated under 3 different scoring and ranking scenarios.
Drink X is always the one that does not get a slot.}

{\normalsize 
\begin{table*}[t]
{\normalsize {\small \vspace{-0cm}
\begin{equation*}
\begin{tabular}{c|c|c|c|c|c|c|c|c|c}
\multicolumn{10}{c}{Table 2: Bids and quality scores for the key word
\textquotedblleft soda drink", senario 1.} \\ \hline
Advertiser & $e_{i}$ & $b_{i}$ & $e_{i}\times b_{i}$ & Rank & $p_{i,j}$ & $%
v_{i}$ & $x_{i,j}$ & $\left( v_{i}-p_{i,j}\right) \times x_{i,j}$ & $%
p_{i,j}\times x_{i,j}$ \\ \hline\hline
Coke & 70 & \$0.05 & 3.5 & 1 & \$0.03 & \$0.10 & 70 & \$4.9 & \$2.1 \\ \hline
Pepsi & 30 & \$0.07 & 2.1 & 2 & \$0.067 & \$0.10 & 30 & \$1.0 & \$2.0 \\
\hline
Dr. Pepper & 20 & \$0.10 & 2.0 & 3 & \$0.07 & \$0.10 & 20 & \$0.60 & \$1.4
\\ \hline
Drink X & 20 & \$0.07 & 1.4 & 4 & \$0 & \$0.07 & 0 & \$0 & \$0 \\ \hline
\end{tabular}%
\end{equation*}
\vspace{-0cm}}  }
\end{table*}
}

{\normalsize 
\begin{table*}[t]
{\normalsize {\small \vspace{-0cm}
\begin{equation*}
\begin{tabular}{c|c|c|c|c|c|c|c|c|c}
\multicolumn{10}{c}{Table 3: Bids and quality scores for the key word
\textquotedblleft soda drink", senario 2.} \\ \hline
Advertiser & $e_{i}$ & $b_{i}$ & $e_{i}\times b_{i}$ & Rank & $p_{i,j}$ & $%
v_{i}$ & $x_{i,j}$ & $\left( v_{i}-p_{i,j}\right) \times x_{i,j}$ & $%
p_{i,j}\times x_{i,j}$ \\ \hline\hline
Pepsi & 50 & \$0.07 & 3.5 & 1 & \$0.05 & \$0.10 & 50 & \$2.5 & \$2.5 \\
\hline
Coke & 50 & \$0.05 & 2.5 & 2 & \$0.04 & \$0.10 & 50 & \$3.0 & \$2.0 \\ \hline
Dr. Pepper & 20 & \$0.10 & 2.0 & 3 & \$0.07 & \$0.10 & 20 & \$0.6 & \$1.4 \\
\hline
Drink X & 20 & \$0.07 & 1.4 & 4 & \$0 & \$0.07 & 0 & \$0 & \$0 \\ \hline
\end{tabular}%
\end{equation*}
\vspace{-0cm}}  }
\end{table*}
}

{\normalsize 
\begin{table*}[t]
{\normalsize {\small \vspace{-0cm}
\begin{equation*}
\begin{tabular}{c|c|c|c|c|c|c|c|c|c}
\multicolumn{10}{c}{Table 4: Bids and quality scores for the key word
\textquotedblleft soda drink", senario 3.} \\ \hline
Advertiser & $e_{i}$ & $b_{i}$ & $e_{i}\times b_{i}$ & Rank & $p_{i,j}$ & $%
v_{i}$ & $x_{i,j}$ & $\left( v_{i}-p_{i,j}\right) \times x_{i,j}$ & $%
p_{i,j}\times x_{i,j}$ \\ \hline\hline
Pepsi & 40 & \$0.08 & 3.2 & 1 & \$0.06125 & \$0.10 & 50 & \$1.875 & \$3.125
\\ \hline
Coke & 50 & \$0.05 & 2.5 & 2 & \$0.04 & \$0.10 & 50 & \$3.0 & \$2.0 \\ \hline
Dr. Pepper & 20 & \$0.10 & 2.0 & 3 & \$0.07 & \$0.10 & 20 & \$0.6 & \$1.4 \\
\hline
Drink X & 20 & \$0.07 & 1.4 & 4 & \$0 & \$0.07 & 0 & \$0 & \$0 \\ \hline
\end{tabular}%
\end{equation*}
\vspace{-0cm}}  }
\end{table*}
In scenario 1, Coke, Pepsi and Dr. Pepper are enjoying surpluses of \$4.9,
\$1.0 and \$0.60 per hour respectively and generating \$2.1, \$2.0 and \$1.4
per hour for the search engine. In such a setting, Dr. Pepper does not have
an incentive to move up in the ranking or otherwise it needs to pay at least
\$0.105 per click which is greater than its value of \$0.10 per click. Dr.
Pepper does not have an incentive to move down either since that will make
its surplus 0. Coke and Pepsi would also prefer their current positions. To
see this, if Coke moves down one slot, it needs to pay at least \$0.0286 and
receives 50 clicks per hour. This ends up with a surplus of \$3.5714, which
is less than the \$4.9 surplus it is making now. Similarly, if Pepsi moves
down one slot, its surplus becomes \$0.8 and if Pepsi moves up on slot, its
surplus is -\$0.8333, both are less than its current surplus \$1.0. Finally,
Drink X does not want to move up one slot because it has to pay \$0.10 which
will make its current 0 surplus being negative. Therefore, scenario 1 is an
NE. Further more, the quality scores for Coke, Pepsi \ and Dr. Pepper
perfectly predict the actual click-through rates after the ranking and
reflect the fact that Coke has a relatively higher \textquotedblleft ad
quality" than Pepsi. However, this \textquotedblleft full mark" scoring
generates the lowest revenue among the 3 scenarios. }

{\normalsize In scenario 2, quality scores for Coke and Pepsi are changed to
be the same of 50 with the remaining advertisers' scores unchanged. This
change results in the swap of ranking for Coke and Pepsi compared with
scenario 1. It is easy to check scenario 2 also represents an NE in
advertisers bidding and that bidding profile is exactly the same as in
scenario 1, but with \$0.4 more revenue generated for the search engine.
Here, quality scores also perfectly predict the actual click-through rates
but do not show the fact that Coke has a better \textquotedblleft ad
quality" than Pepsi. }

{\normalsize Tables 2-3 together show that if the purpose of scoring is to
accurately predict the click-through rates and thus the expected revenue
from each advertiser, scores do not necessarily need to reflect the true
relative \textquotedblleft ad qualities". }

{\normalsize Scenario 3 generates the highest revenue for the search engine
among the 3 scoring profiles and forms another NE with an increased bid of
Pepsi from \$0.07 to \$0.08. This increase in bid is a result of the
decreased ad-quality score for Pepsi compared with scenario 2. It is
interesting to note that increasing the bid from \$0.07 to \$0.08 does not
increase Pepsi's pay per click rate but eliminates Coke's incentive of
moving up one slot. In this scenario, the scores reflect the relative
\textquotedblleft ad qualities" for Coke and Pepsi but do not agree with the
actual click-through rates. }

{\normalsize The above described simple example shows many interesting
observations about ad-quality scoring and ranking. }

1. {\normalsize Rankings can be changed by manipulating the ad-quality
scores $e_{i}$ for corresponding advertisers to serve the purpose of
increasing the search engine's revenue. This can be done with (i.e.,
scenario 3) or without (i.e., scenario 2) changing the bidding profile of
the advertisers. }

2. {\normalsize The scoring profile that gives the search engine more
revenue does not necessarily need to reflect better the relative
\textquotedblleft ad-qualities" of advertisers or the actual click-through
rate after the ranking. }

3. {\normalsize Changes in the ad-quality scores can result in the
advertisers' bids moving from one NE to another. Any increase in the search
engine's revenue comes from the decrease in the advertisers' total
surpluses. }

\section{\protect\normalsize Scoring Strategy in Complete Information Setting%
}

{\normalsize The nature of the ad-quality score and the dynamic interactions
between the search engine's scoring and the advertisers bidding make search
engine revenue maximization far from a straightforward business. Under the
same scoring profile, there might be multiple NE bidding sets and multiple
resulting rankings. The existence of multiple equilibria also adds to the
difficulty in reasoning about the search engine revenue generated by the ad
auction, since it depends on which equilibrium (potentially from among many)
is selected by the advertisers. In light of this, we choose to start our
analysis with the socially optimal ranking, which is independent with the
search engine and advertisers' behavior. Besides, the maximum social surplus
obtained under the socially optimal ranking serves naturally as an upper
bound for the search engine's maximum achievable revenue. }

{\normalsize Let us add fake slots with click-through rates $x_{\cdot ,j}=0$
for $j=S+1,...,N$ to make the total number of slots match the total number
of advertisers. Let $\sigma $ be a one-to-one permutation function from the
advertisers' index set $\left\{ 1,...,N\right\} $ to the slots index set $%
\left\{ 1,...,N\right\} $, so that for advertiser $i$, $\sigma \left(
i\right) $ gives its rank. Let $\Pi _{i}\left( \sigma \right) $ be the
surplus for advertiser $i$ with respect to ranking $\sigma $ and $\Pi
_{ad}\left( \sigma \right) =\sum_{i=1}^{N}\Pi _{i}\left( \sigma \right) $ be
the corresponding total surpluses for all advertisers. For any permutation $%
\sigma $, the social surplus $\Pi _{social}\left( \sigma \right) $ is the
sum of total advertisers' surpluses $\Pi _{ad}\left( \sigma \right) $ and
the search engine's surplus $\Pi _{se}\left( \sigma \right) $, where
\begin{eqnarray*}
\Pi _{social}\left( \sigma \right) &=&\Pi _{ad}\left( \sigma \right) +\Pi
_{se}\left( \sigma \right) \\
&=&\sum_{i=1}^{N}\left( v_{i}-p_{i,\sigma \left( i\right) }\right)
x_{i,\sigma \left( i\right) }+\sum_{i=1}^{N}p_{i,\sigma \left( i\right)
}x_{i,\sigma \left( i\right) } \\
&=&\sum_{i=1}^{N}v_{i}x_{i,\sigma \left( i\right) }
\end{eqnarray*}%
}

{\normalsize \bigskip }

{\normalsize \textbf{Definition 2}. A socially optimal ranking $\sigma
^{\ast }$ is the permutation that maximize the social surplus, where%
\begin{equation*}
\sigma ^{\ast }=\arg \max_{\sigma }\Pi _{social}\left( \sigma \right) =\arg
\max_{\sigma }\sum_{i=1}^{N}v_{i}x_{i,\sigma \left( i\right) }
\end{equation*}%
}

{\normalsize Before we move on to analyze the scoring strategy for the
search engine, }we assume the goal for any advertiser is surplus
maximization.

\bigskip

{\normalsize \textbf{Definition 3}. Given the values of the advertisers $%
\left( v_{1},...,v_{N}\right) $, an equalizing\emph{\ scoring profile} for
the search engine is defined as a scoring profile $e^{\ast }=\left(
e_{1}^{\ast },...,e_{N}^{\ast }\right) $ such that $e_{i}^{\ast
}v_{i}=e_{i^{\prime }}^{\ast }v_{i^{\prime }}$ for all $i\neq i^{\prime }$. }

{\normalsize By definition, an equalizing scoring profile always exists,
i.e., $e_{i}^{\ast }=1/v_{i}$ for all $i$. If $e^{\ast }$ is an equalizing
scoring profile, for any $k>0$, $ke^{\ast }$ is also an equalizing scoring
profile. The following Theorem 1 shows that where there are more advertisers
than slots with }$N>S$, {\normalsize under an equalizing scoring profile,
the set of truthful bids forms a Nash equilibrium. Furthermore, in such an
NE, the search engine extracts all the surpluses from the bidders making the
search engine's surplus equal the total social surplus. }

{\normalsize \bigskip }

{\normalsize \textbf{Theorem 1}. When there are more advertisers than slots,
}$N>S$, under any equalizing scoring profile $e^{\ast }$, the set of
truthful bids $b^{\ast }=\left( v_{1},...,v_{N}\right) $ is an NE. The
surplus for each advertiser $\Pi _{i}=0$ and the search engine's surplus $%
\Pi _{se}=\Pi _{social}$. The resulting ranking can be any permutation $%
\sigma $ of the advertisers.

\textbf{Proof.} Given the set of truthful bids $b^{\ast }=\left(
v_{1},...,v_{N}\right) $, we have $\ b_{i}=v_{i}$ and thus the price
advertiser $i$ paying is $p_{i}=v_{1}e_{1}/e_{i}=v_{i}$, for all $i=1,...,N$%
. Without loss of generality, suppose advertiser $1$ bids above its value $%
v_{1}$. In this case, it is assigned the top slot and the price for the top
slot changes to $p_{1,1}=e_{i}v_{i}/e_{1}=v_{1}$ which leads to no positive
surplus. Hence, this deviation is not strictly profitable. If an advertiser
bids below its value then it loses a slot if it had been allocated one since
$N>S$, or otherwise, it continues to have no slot allocated. Again, no
positive surplus. This shows the set of bids $b^{\ast }=\left(
v_{1},...,v_{N}\right) $ is an NE.

Each advertiser's surplus $\Pi _{i}\left( \sigma \right) =\left(
v_{i}-p_{i,\sigma \left( i\right) }\right) x_{i,\sigma \left( i\right) }=0$.
Since this is true for any permutation $\sigma $, any permutation $\sigma $
can be the resulting ranking from the set of truthful bids $b^{\ast }=\left(
v_{1},...,v_{N}\right) $. \ \ \ \ \ \ \ \ \ \ \ \ \ \ \ \ \ \ \ \ \ \ \ \ \
\ \ \ \ \ \ \ \ \ \ \ \ \ \ \ \ \ \ \ \ \ \ \ \ \ \ \ \ \ \ \ \ \ \ \ \ \ \
\ \ \ \ \ $\blacksquare $

\bigskip

{\normalsize In the proof of Theorem 1, it shows that under an equalizing
scoring profile $e^{\ast }$, the set of truthful bids forms an NE but does
not indicate whether this NE is unique. It is easy to check that a set of
bids where one advertiser bids above its value while all other advertisers
bids their values also represents an NE and therefore, the truthful bidding
set is not the unique NE that an equalizing scoring profile }$e^{\ast }$
will induce.

{\normalsize In Theorem 2, we show that when }$N>S$, {\normalsize under the
equalizing scoring profile $e^{\ast }$ and the set of truthful bids $b^{\ast
}$, among the $N!$ possible NE permutations, the one that maximize the
search engine's surplus is the socially optimal ranking $\sigma ^{\ast }$. }

{\normalsize \bigskip }

{\normalsize \textbf{Theorem 2}. When }$N>S$, under the equalizing scoring
profile $e^{\ast }$, the set of truthful bids $b^{\ast }$ and the socially
optimal ranking $\sigma ^{\ast }=\arg \max_{\sigma }\sum_{i=1}^{N}$ $%
v_{i}x_{i,\sigma \left( i\right) }$ together form an NE where each
advertiser has $0$ surplus and the search engine has its maximum revenue
equal to the maximum social surplus $\Pi _{se}^{\ast }=\sum_{i=1}^{N}$ $%
v_{i}x_{i,\sigma ^{\ast }\left( i\right) }$.

\textbf{Proof}. The results follow directly from Theorem 1. \ \ \ \ \ \ \ \
\ $\blacksquare $

{\normalsize \bigskip }

{\normalsize It is commonly assumed that the expected click-through rate of
advertiser $i$ in slot $j$\ can be written as the product of an ad-specific
factor $q_{i}$ and a position-specific factor $s_{j}$, specifically $%
x_{ij}=q_{i}s_{j}$. While this product form is not generally true in
practice, it makes the ad-specific factor and the position specific factor
separable and usually leads to particularly simple analytical results.}

{\normalsize \bigskip }

{\normalsize \textbf{Assumption 1}. The expected click-through rate
advertiser $i$ gets in slot $j$ can be expressed in a product form $%
x_{ij}=q_{i}s_{j}$, where $q_{i}\geq 0$ is an ad-specific factor and $%
s_{j}\geq 0$ is a position-specific factor. Furthermore, we assume the
position-specific factors are in a descending order with respect to the
ranking of the slots, that is $s_{j}>s_{j^{\prime }}$, for all $j<j^{\prime
} $. }

{\normalsize \bigskip }

{\normalsize \textbf{Theorem 3}. Under Assumption 2, a permutation $\sigma $
is the socially optimal ranking iff for any adjacently ranked advertisers $%
\left( i,i^{\prime }\right) $, where $\sigma \left( i\right) =\sigma \left(
i^{\prime }\right) -1$, $v_{i}x_{i,\sigma \left( i\right) }+v_{i^{\prime
}}x_{i^{\prime },\sigma \left( i^{\prime }\right) }\geq v_{i}x_{i,\sigma
\left( i^{\prime }\right) }+v_{i^{\prime }}x_{_{i,}\sigma \left( i\right) }.$
}

{\normalsize \textbf{Proof}.\textbf{\ }$``\Rightarrow "$. Obvious by the
definition of social optimality. }

{\normalsize $``\Leftarrow "$. Substituting $\sigma \left( i^{\prime
}\right) =\sigma \left( i\right) +1$, $x_{i},_{\sigma \left( i\right)
}=q_{i}s_{\sigma \left( i\right) }$ and $x_{i^{\prime }},_{\sigma \left(
i^{\prime }\right) }=q_{i^{\prime }}s_{\sigma \left( i^{\prime }\right) }$
into $v_{i}x_{i,\sigma \left( i\right) }+v_{i^{\prime }}x_{i^{\prime
},\sigma \left( i^{\prime }\right) }\geq v_{i}x_{i,\sigma \left( i^{\prime
}\right) }+v_{i^{\prime }}x_{_{i,}\sigma \left( i\right) }$, we have%
\begin{eqnarray*}
v_{i}q_{i}s_{\sigma \left( i\right) }+v_{i^{\prime }}q_{i}^{{}}s_{\sigma
\left( i\right) +1} &\geq &v_{i}q_{i}s_{\sigma \left( i\right)
+1}+v_{i^{^{\prime }}}q_{i^{\prime }}^{{}}s_{\sigma \left( i\right) } \\
v_{i}q_{i}\left( s_{\sigma \left( i\right) }-s_{\sigma \left( i\right)
+1}\right) &\geq &v_{i^{\prime }}q_{i^{\prime }}\left( s_{\sigma \left(
i\right) }-s_{\sigma \left( i\right) +1}\right)
\end{eqnarray*}%
By the descending property of the position-specific factors, we have $%
s_{\sigma \left( i\right) }-s_{\sigma \left( i\right) +1}>0$, which gives $%
v_{i}q_{i}\geq v_{i^{\prime }}q_{i^{\prime }}$. Since this is true for all
adjacently ranked advertisers $\left( i,i^{\prime }\right) $, it means the
permutation $\sigma $ is a ranking based on the descending order of $%
v_{i}q_{i}$. Without loss of generality, we re-index the advertisers such
that for all $i<i^{\prime }$, $v_{i}q_{i}\geq v_{i^{\prime }}q_{i^{\prime }}$%
. Then $\sigma \left( i\right) =i$, meaning advertiser $i$ is assigned to
slot $i$ for all $i=1,...,N$. }

{\normalsize Now let us assume the socially optimal ranking is $\sigma
^{\ast }\neq \sigma $. Then there is at lease one $i$, such that $\sigma
^{\ast }(i)\neq i$. Let $k=\min \left\{ i|\sigma ^{\ast }(i)\neq i\right\} $
and $k^{\prime }$ be the index of the advertiser such that $\sigma ^{\ast
}\left( k^{\prime }\right) =k$. Define another permutation $\sigma ^{\prime
} $ that only differs with $\sigma ^{\ast }$ by swapping the ranks of
advertisers $k$ and $k^{\prime }$. Specifically, $\sigma ^{\prime }\left(
k\right) =\sigma ^{\ast }\left( k^{\prime }\right) $, $\sigma ^{\prime
}\left( k^{\prime }\right) =\sigma ^{\ast }\left( k\right) $\ and $\sigma
^{\prime }\left( i\right) =\sigma ^{\ast }\left( i\right) $, for\ all $i\neq
k,k^{\prime }$. The social surplus under the new permutation $\sigma
^{\prime }$ is%
\begin{eqnarray*}
&&\Pi _{social}\left( \sigma ^{\prime }\right) \\
&=&\Pi _{social}\left( \sigma ^{\ast }\right) -v_{k}x_{k,\sigma ^{\ast
}\left( k\right) }-v_{k^{\prime }}x_{k^{\prime },\sigma ^{\ast }\left(
k^{\prime }\right) } \\
&&+v_{k}x_{k,\sigma ^{\prime }\left( k\right) }+v_{k^{\prime }}x_{k^{\prime
},\sigma ^{\prime }\left( k^{\prime }\right) } \\
&=&\Pi _{social}\left( \sigma ^{\ast }\right) +v_{k}q_{k^{{}}}\left(
s_{\sigma ^{\prime }\left( k\right) }-s_{\sigma ^{\ast }\left( k\right)
}\right) \\
&&-v_{k^{\prime }}q_{k^{\prime }}\left( s_{\sigma ^{\ast }\left( k^{\prime
}\right) }-s_{\sigma ^{\prime }\left( k^{\prime }\right) }\right) \\
&=&\Pi _{social}\left( \sigma ^{\ast }\right) +\left(
v_{k}q_{k^{{}}}-v_{k^{\prime }}q_{k^{\prime }}\right) \left( s_{\sigma
^{\ast }\left( k^{\prime }\right) }-s_{\sigma ^{\ast }\left( k\right)
}\right) \\
&\geq &\Pi _{social}\left( \sigma ^{\ast }\right)
\end{eqnarray*}%
}

{\normalsize The last inequality follows the fact that $k$ is the first
advertiser that does not satisfy $\sigma ^{\ast }(i)=i$ and therefore $%
v_{k}q_{k^{{}}}-v_{k^{\prime }}q_{k^{\prime }}\geq 0$ and $s_{\sigma ^{\ast
}\left( k^{\prime }\right) }-s_{\sigma ^{\ast }\left( k\right) }>0$. Since $%
\sigma ^{\ast }$ is a socially optimal ranking, $\sigma ^{\prime }$ is also
a socially optimal ranking. By sequentially swapping all the out-of-place $i$%
s, we will finally obtain the permutation $\sigma \left( i\right) =i$ and $%
\Pi _{social}\left( \sigma \right) \geq \Pi _{social}\left( \sigma ^{\ast
}\right) $. Therefore, $\sigma \left( i\right) =i$ is the socially optimal
permutation. \ \ \ \ \ \ \ \ \ \ \ \ \ \ \ \ \ \ \ \ \ \ \ \ \ \ \ \ \ \ \ \
\ \ \ \ \ \ \ \ \ \ \ \ \ \ \ \ \ \ \ \ \ \ \ \ \ \ \ \ \ \ \ \ \ \ \ \ \ \
\ $\blacksquare $ }

{\normalsize \bigskip }

{\normalsize The proof of Theorem 3 shows a constructive way of finding the
socially optimal ranking under the equalizing scoring profile by
sequentially swapping neighboring ads seeking improvement of the total
revenue until no improvement can be made. The is an example where local
search leads to global optimum and the search algorithm costs no more than a
bubble sort! }

{\normalsize Theorems 1-3 all together serve as the theoretical foundation
for constructing the adaptive online ad-auction scoring algorithm shown in
the next section. }

\section{\protect\normalsize An adaptive online ad-auction scoring algorithm}

{\normalsize In this section, we develop an adaptive online ad-auction
scoring algorithm to maximize the search engine's revenue in the incomplete
information case where the values of the advertisers are not known to the
search engine. }

\subsection{\protect\normalsize Arguments on Advertisers Behavior}

{\normalsize Before we move on to the algorithm construction, we make two
arguments of the advertiser behavior based on the surplus maximization
assumption for each advertiser}

\textbf{Argument 1}{\normalsize . \emph{An advertiser will not bid above its
true value}.}

{\normalsize We have shown in the complete information setting, at least
with the equalizing scoring profile, one advertiser bids above its value and
other bids their values is an NE and therefore overbidding is not an
infeasible strategy. However, in practice, the search engine does not
provide an advertiser with its own or others \textquotedblleft ad-quality"
scores or others bids. While over time, an advertiser is likely to learn all
the information to identify that itself is in an equilibrium, try bidding
over its value always has a chance of being penalized with negative surplus.}

\textbf{Argument 2}{\normalsize . \emph{If being left out without having a
slot, an advertiser will bid its true value}.}

{\normalsize By surplus maximization, the goal for any advertiser joining
the auction is to get a slot and maximize its surplus. Since the bid is an
advertiser's only strategy to get itself a slot, when keeps missing a slot,
it will at least try to express its maximum willingness to pay to see
whether that can get it a slot.}

{\normalsize Though we are not proving the above two arguments, they are not
unreasonable in practice, especially when the scoring mechanism is kept
totally secret by the search engine. In what follows, we propose an adaptive
online ad auction scoring algorithm for the search engine revenue
maximization taking advantage of the above two arguments. Though the results
we get may be too ideal in reality, it illustrates a possible approach to
reveal advertisers' values and shows insights into what will happen when the
search engine have a way to accurately estimate advertisers' values.}

\subsection{\protect\normalsize Algorithm Construction}

{\normalsize Theorems 1-3 show that in complete information setting, how a
search engine may maximize its revenue, 1) apply an equalizing scoring
profile; 2) induce truthful bidding; 3) find the socially optimal ranking.}

{\normalsize To apply the equalizing scoring strategy, the search engine
must have a mechanism to find out first the values of the advertisers. Once
the values are found, an optimization procedure is needed to search for the
optimal ranking to maximize the revenue. Accordingly, we have in our scoring
algorithm a \textquotedblleft value revealing" module and an
\textquotedblleft optimal rank searching" module to implement these two
functionalities simultaneously. The idea is based on the following. }

{\normalsize 1. We divide the set of advertisers into two sets, one set $U$
contains all the advertisers whose values are not known yet, the other set $%
V=\left\{ 1,...,N\right\} \backslash U$ contains those whose values are
revealed already. By Argument 2, after the first NE is reached, $V\neq \phi $%
. }

{\normalsize 2. The value revealing module applies equalizing scoring
strategy to all the advertisers in $V$ to make their $e_{i}\times v_{i}$
equal (and therefore everyone is forced to bid their true value) and then
gradually increases $e_{i}s$ (while making sure all $e_{i}\times v_{i}$
increase the same amount) to push out surpluses from the advertisers in set $%
U$. Since the ad-quality scores for advertisers in $U$ are not increased, an
advertiser in $U$ will ultimately be forced to lose its slot and reveal its
value to the search engine. This is when the advertiser leaves the set $U $
and enters $V$. }

{\normalsize 3. Once a new advertiser $i$ enters $V$, the optimal rank
searching module applies sequential swaps to the new member $i$ and its
neighboring advertiser $j$ by making a small $\varepsilon >0$ difference
between $e_{i}b_{i}$ and $e_{j}b_{j}$. The swap will keep going until no
increase in the search engine's revenue can be obtained. By Theorem 3, when
advertisers enter the set $V$ gradually, the optimal rank searching module
makes sure all the advertisers in the set $V$ are ordered in a way that is
consistent with the social optimal ranking. }

{\normalsize 4. Any time an advertiser's score is changed, the algorithm
will let all the advertisers settle into a new NE. By the way how the value
revealing module works, for any $i\in V$ and $j\in U$, we always have $%
e_{i}v_{i}<e_{j}v_{j}$, and the auction rules will make sure the permutation
$\sigma $ resulted from NE has $\sigma \left( i\right) >\sigma \left(
j\right) $. }

{\normalsize 5. The algorithm continues until the set $U$ is emptied. This
is when all the advertisers' values are revealed and resulting permutation $%
\bar{\sigma} $ is the socially optimal ranking $\sigma ^{\ast }$. }

{\normalsize
\begin{figure}[tbh]
{\normalsize \centering \includegraphics[width=%
\linewidth]{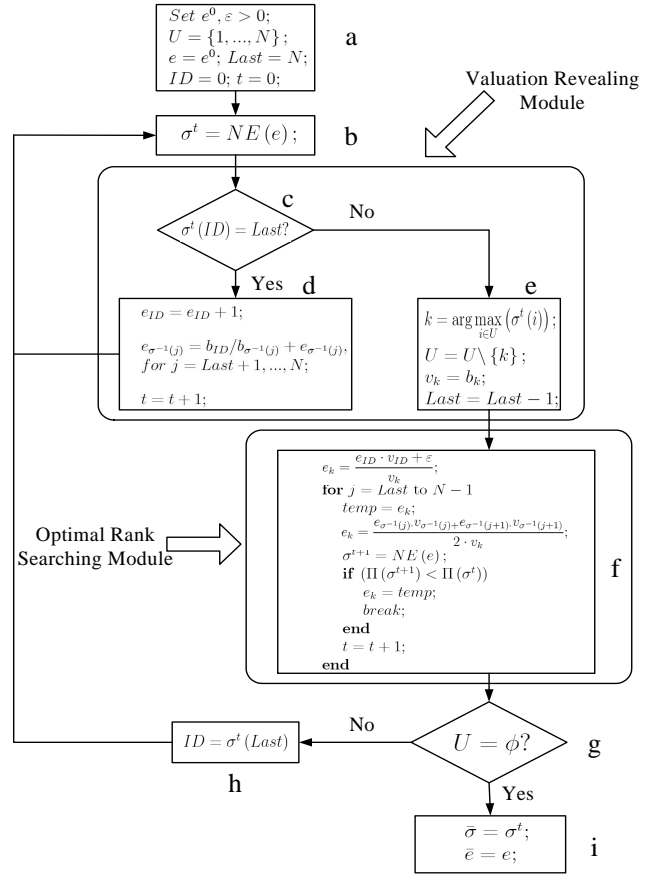}  }
\caption{ Flow chart for the adaptive scoring algorithm }
\label{figFlowChart}
\end{figure}
}

{\normalsize 
}

{\normalsize Figure \ref{figFlowChart} illustrates the flow chart with
pseudo code for the adaptive ad scoring algorithm. $\sigma ^{-1}$ is the
inverse function of $\sigma $ so that given a slot index $j\,$, $\sigma
^{-1}\left( j\right) $ returns the index of the advertisers occupying slot $%
j $. }

{\normalsize $ID$ is the index of the advertiser in the set $V$ that has the
highest rank, where $ID=\arg \min_{i}\left\{ \sigma \left( i\right) |i\in
V\right\} $. $Last$ is the index of the slot that equals the total number of
advertisers in set $U$ plus 1, where $Last=\left\vert U\right\vert +1$. $%
NE\left( \cdot \right) $ is a function that takes the input of a scoring
profile $e=\left( e_{1},...,e_{N}\right) $, simulates the Nash equilibrium
of advertisers' bids and then returns a resulting permutation $\sigma $. $t$
keeps track of the total number of score adjustments needed for the
algorithm. }

{\normalsize Under Assumption 1 and by Theorems 1-3, the final ranking $\bar{%
\sigma}$ obtained by the proposed algorithm is the socially optimal ranking $%
\sigma ^{\ast }$. The total search engine revenue $\Pi _{se}\rightarrow $ $%
\Pi _{social}\left( \sigma ^{\ast }\right) $ as $\varepsilon \rightarrow 0$.
}

\section{\protect\normalsize Empirical analysis: implementation and results}

{\normalsize We test the proposed algorithm under the 8-slot sponsored ad
setting used by Google, Microsoft (Bing.com) and Yahoo! with 9 advertisers.
Given 8 slots, 9 advertisers, there will be $9!\approx 360,000$
click-through rate distribution vectors. We first show that our algorithm
finds the social optimal ranking when the system parameters $v$, $q$ and $s$
take static values. We then show that in a dynamic environment, our
algorithm is able to adapt to the changes of these parameters in a timely
fashion. }

{\normalsize Finally, we relax the assumption that the click-through rate
takes a product form of a ad-specific factor and a position-specific factor
and make the click-through rate also dependent on the relative positions
among the advertisers. Under this new model for click-through rate, our
algorithm can not claim optimality. However, we show by simulation that the
maximum revenue for the search engine obtained by our algorithm is close to
the optimal search engine revenue. }

\subsection{\protect\normalsize 8-Slot Auction with Static System Parameters}

{\normalsize We set the user value vector as $v=$}(19, 8, 7, 6, 5, 4, 13,
12, 1),{\normalsize \ the position-specific factor vector as $s=$(65, 50,
40, 36, 30, 18, 12, 10, 0) and the ad-specific factor vector as $q=$(35, 45,
35, 20, 50, 20, 10, 70, 5). (we need these two vectors to generate the
actual click through rates $x$ for different permutations). The socially
optimal ranking for this scenario is $\sigma ^{\ast }=$(2, 3, 5, 7, 4, 8, 6,
1, 9) and the optimal social surplus $\Pi _{social}(\sigma ^{\ast })=123,180$%
. }

{\normalsize We set the initial scores for all the advertisers to be $100$,
i.e., $e^{0}=\left( 100,...,100\right) $. The algorithm runs $338$
iterations (i.e., it adjusts the ad-quality scores 338 times) before it
stops. We show in Table. 4 the output of the algorithm, where $\bar{b}_{i}$,
$\bar{e}_{i}$, $\bar{p}_{i}$ denote the final bidding, ad-quality score and
price for each advertiser $i$ respectively.
\begin{table}[tbh]
\begin{center}
{\normalsize {\ {\small 
\begin{tabular}{c|c|c|c|c}
\multicolumn{5}{c}{Table 4: Output of the algorithm for static case.} \\
\hline
Advertiser~$i$ & rank~$\bar{\sigma}(i)$ & $\bar{b}_i\times \bar{e}_i$ & $%
\bar{b}_i$ & $\bar{p}_i$ \\ \hline\hline
1 & 2 & 1909 & 19 & 18.94 \\ \hline
2 & 3 & 1903 & 8 & 7.97 \\ \hline
3 & 5 & 1893.7247 & 7 & 6.997 \\ \hline
4 & 7 & 1892.5 & 6 & 5.992 \\ \hline
5 & 4 & 1895 & 5 & 4.997 \\ \hline
6 & 8 & 1890 & 4 & 3.987 \\ \hline
7 & 6 & 1893.125 & 13 & 12.995 \\ \hline
8 & 1 & 1915 & 12 & 11.962 \\ \hline
9 & 9 & 1886 & 1 & 0 \\ \hline
\end{tabular}
} }  }
\end{center}
\par
{\normalsize \vspace{-0.1cm}
}
\end{table}
}

{\normalsize We also obtain the final revenue for the search engine to be $%
\Pi _{se}(\bar{\sigma})=122,032.85$, which is $99.07\%$ of the social
optimal revenue $\Pi _{social}({\sigma }^{\ast })=123,180$. }

{\normalsize Our first observation is that the output ranking $\bar{\sigma}$
equals the social optimal ranking $\sigma ^{\ast }$. The second observation
is that the price $\bar{p}_{i}$ each advertiser $i$ pays almost equal to its
individual value $v_{i}$, which means each advertiser has almost no surplus
left. Combining the previous two observations, we conclude that the search
engine gets almost all the surplus. We see that this is indeed the case when
we compare $\Pi _{social}(\sigma ^{\ast })$ and $\Pi _{se}(\bar{\sigma})$.
The third observation is that $e_{i}\times b_{i}$ is ranked the same as the
advertiser ranking, with the advertiser who ranks higher gets larger $%
e_{i}\times b_{i}$. This property ensures that the rule of the ranking is
observed, i.e., the search engines should order the advertisers by $%
b_{i}\times e_{i}$. }

{\normalsize Fig.\ref{figEBEvolution} shows the evolution of the values $%
e_{i}^{t}\times b_{i}^{t}$ for all advertisers and illustrates how our
algorithm works. In this example, advertiser 9 has both the lowest value and
ad-specific factor, therefore remains ranked last throughout the process.\
Advertiser $6$ has the next lowest value and is forced by the value
revealing module to bid its value at approximately $t=150$. This is when the
two trajectories $e_{6}^{t}\times b_{6}^{t}$ and $e_{9}^{t}\times b_{9}^{t}$
meet and advertiser 6 enters the set $V$ from then on. Then advertiser $5$,
who has the third lowest value, is forced to bid its value at approximately $%
t=175$. This process goes on until at time $t=334$, when advertiser $1$, who
has the highest value, is forced to bid its value. At this point, the $%
b_{i}^{t}\times e_{i}^{t}$ becomes almost equal for everyone. }

{\normalsize We notice that advertisers join the set $V$ in the order of
their values. However, the final rank $\bar{\sigma}$ is in the order of $%
q_{i}\times v_{i}$ by the work of the optimal rank searching module. The
small $\varepsilon $ difference in $e_{i}\times b_{i}$ plays the role of
maintaining the relative positions of the advertisers in $V$ according to
the social optimal ranking. Because of this $\varepsilon $ difference in $%
e_{i}\times b_{i}$, the price each advertiser pays is slightly lower than
its true value and therefore making the search engine's revenue a little
less than the total social surplus. This little gap can be view as a cost to
the search engine in order to keep the advertisers ranked in the way it
wants. }

{\normalsize 
}

{\normalsize
\begin{figure}[ht]
{\normalsize {\includegraphics[width=\linewidth]{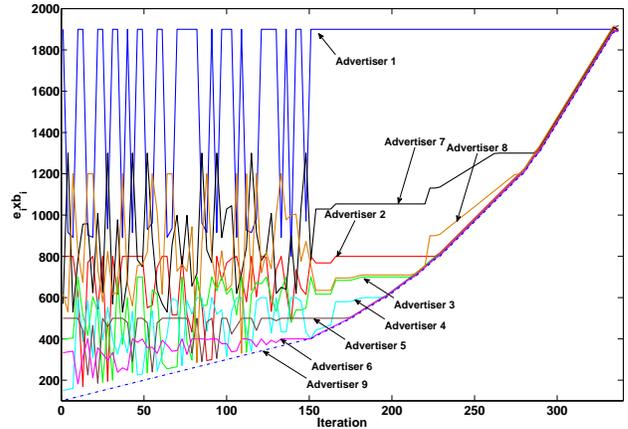} }
\vspace{-.2cm} \vspace*{-.3cm}  }
\caption{ Evolution of $e_i\times b_i$ produced by the algorithm for all
advertisers. }
\label{figEBEvolution}
\end{figure}
}

\subsection{\protect\normalsize 8-Slot Auction with Dynamic System Parameters%
}

{\normalsize In this subsection, we show that once our algorithm has found a
socially optimum ranking, it is able to track small changes in the
individual value, the position-specific factor, or the ad-specific factor in
a timely fashion. }

{\normalsize In order to do so, we assume that these parameters are time
varying, and we use $v^{u}$, $q^{u}$ and $s^{u}$ to denote the value of
these parameters at time instance $u$, respectively. We are interested in
the number of adjustments of the ad-quality scores the search engine needs
to perform before the old optimum profile (optimal ranking and optimal
scores) can adapt to the new system parameters and arrives at a new optimum
profile. }

{\normalsize Now, we generate the parameters at time $u$ by }$v^{u}=\hat{v}%
+\epsilon _{v}^{u}$, $q^{u}=\hat{q}+\epsilon _{q}^{u}$, $s^{u}=\hat{s}%
+\epsilon _{s}^{u}$, {\normalsize where $\hat{v}$, $\hat{q}$ and $\hat{s}$
are the mean values of the parameters; $\epsilon _{v}^{u}$, $\epsilon
_{q}^{u}$ and $\epsilon _{s}^{u}$ are the noisy realization to these
parameters at time $u$; and $v^{u}$, $q^{u}$ and $s^{u}$ are the actual
values these parameters take at time $u$. }

{\normalsize The simulation setting is as follows. In order to compare the
results obtained in the previous subsection, we choose $\hat{v}=$(19, 8, 7,
6, 5, 4, 13, 12, 1), $\hat{q}=$(65, 50, 40, 36, 30, 18, 12, 10, 5), $\hat{s}%
= $(35, 45, 35, 20, 50, 20, 10, 70, 0), which are the same as the
corresponding static parameters in the previous subsection. We set $\epsilon
_{q}\sim \mathcal{N}(0,36)$, $\epsilon _{s}\sim \mathcal{N}(0,25)$ and $%
\epsilon _{v}\sim \mathcal{N}(0,0.1)$. This setting implies
position-specific factors are more volatile than the ad-specific factors and
the values of the advertisers are almost static. }

{\normalsize We run the algorithm 11 times. The initial setting is $v^{1}=%
\hat{v}$, $q^{1}=\hat{q}$, $s^{1}=\hat{s}$, with a randomized initial
ranking and $e_{i}^{0}=100$ for all advertisers. The values of $v$, $q$, $s$
remain the same until the algorithm finds the social optimum profile at time
$u$, then $v^{u+1},$ $q^{u+1}$ and $s^{u+1}$ are perturbed by $\epsilon _{q}$%
, $\epsilon _{s}$ and $\epsilon _{v}$. The algorithm restarts by using the
ranking and ad-quality score profile obtained at time $u$, and adjusts the
scores to adapt to the new values of $v^{u+1}$, $q^{u+1}$ and $s^{u+1}$. }

{\normalsize In Fig. \ref{figRankingTimeVarying}, we show the evolution of
resulting ranking produced by our algorithm\footnote{{\normalsize We check
that at each time instance, the ranking produced by our the algorithm is the
same as the social optimal ranking.}}. }

{\normalsize 
}

{\normalsize
\begin{figure}[ht]
{\normalsize {\includegraphics[width=1%
\linewidth]{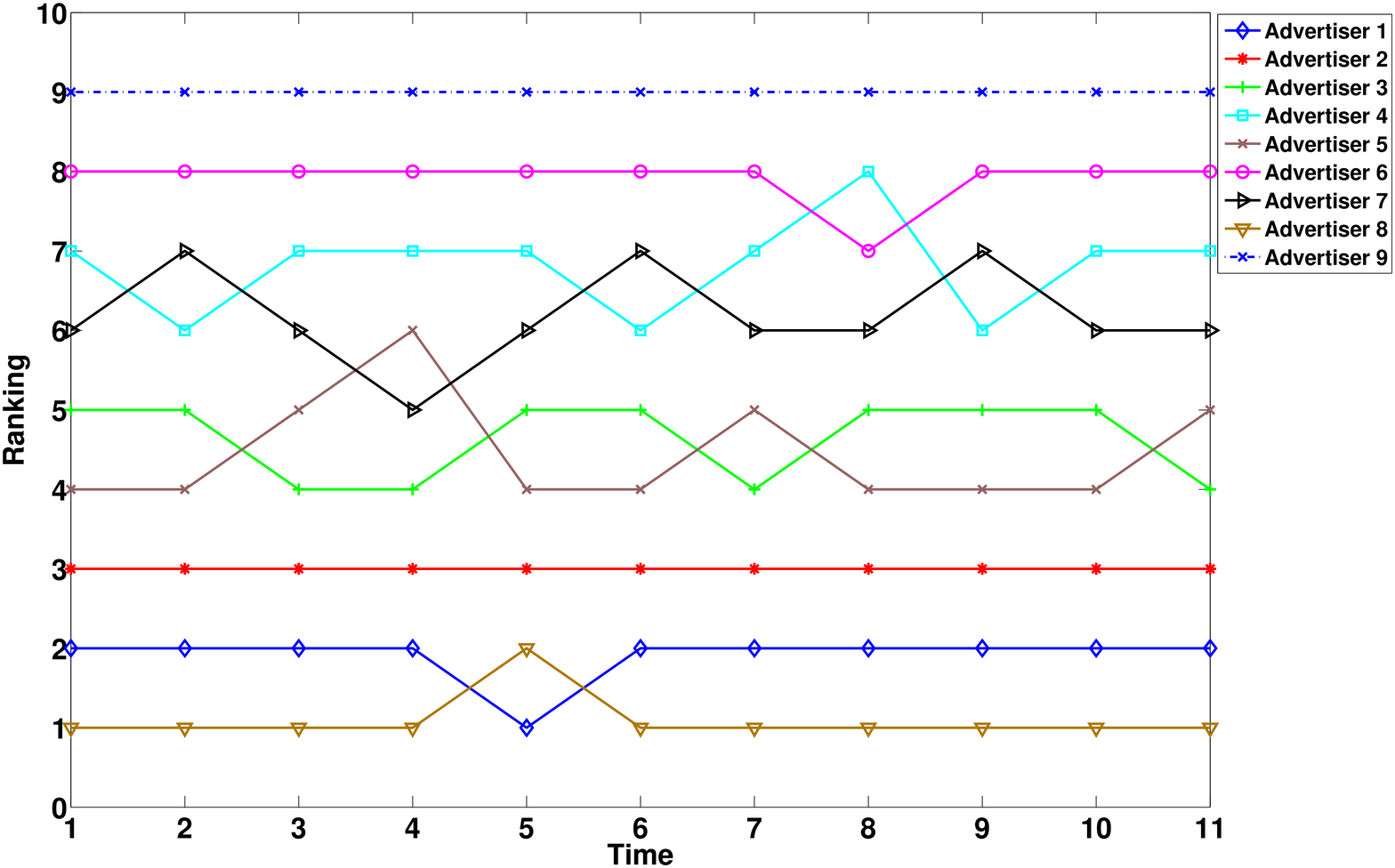} } \vspace{-.2cm} \vspace*{-.3cm%
}  }
\caption{ Evolution of $\bar{\protect\sigma}^u$ produced by the algorithm
for all advertisers. }
\label{figRankingTimeVarying}
\end{figure}
}

{\normalsize In Fig. \ref{figAdjustmentTimeVarying}, we show the number of
adjustments of the ad-quality scores needed for each time instance. We
compare two cases in this graph. Case 1: $\epsilon _{q}\sim \mathcal{N}%
(0,36) $, $\epsilon _{s}\sim \mathcal{N}(0,25)$, $\epsilon _{v}\sim \mathcal{%
N}(0,0.1)$; Case 2: $\epsilon _{q}\sim \mathcal{N}(0,100)$, let $\epsilon
_{s}\sim \mathcal{N}(0,64)$, $\epsilon _{v}\sim \mathcal{N}(0,1)$. It is
clear that the parameters of Case 2 are much more volatile than those of
case 1. We observe that more volatility in the system parameters results in
more adjustments for the algorithm to find the optimal ranking. However, in
both cases, once the algorithm finds the social optimal ranking in time
instance $1$, the adjustments needed for the algorithm to adapt to the new
parameters in the subsequent time instances are small. }

{\normalsize 
}

{\normalsize
\begin{figure}[th]
{\normalsize {\includegraphics[width=%
\linewidth]{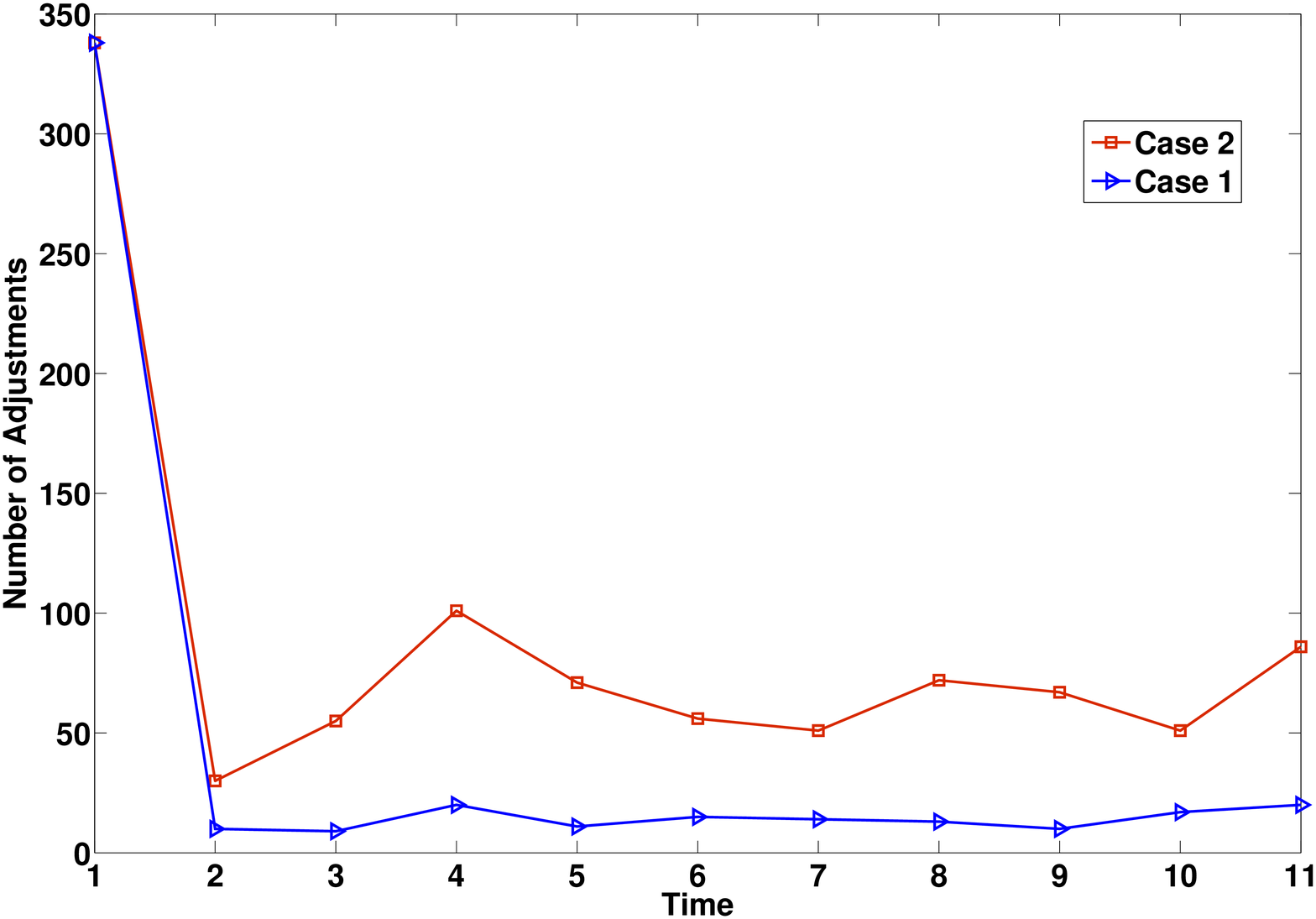} } \vspace{-.4cm} \vspace{%
-.2cm}  }
\caption{ {\protect\small The number of score adjustments needed for each
perturbation of the system parameters.} }
\label{figAdjustmentTimeVarying}
\end{figure}
}

{\normalsize \bigskip }

\subsection{\protect\normalsize 8-Slot Auction with Modified Click-Through
Rate Model}

{\normalsize We first describe a modified click-through rate model that
takes into consideration the effect relative position between direct
competitors on the click-through rates. We assume that among the 9
advertisers, some of them are direct competitors in selling the same type of
product (we call them \textquotedblleft Group 1 Advertisers"), while the
rest of them are not in direct competition with any of the other advertisers
(we call them \textquotedblleft Group 2 Advertisers"). An example for this
situation in reality is that when we search "EOS Camera" on www.goole.com, 6
of the sponsored ads are related to the web sites that sells new Canon EOS
cameras; 1 of them sells refurbished EOS cameras, and 1 of them sells
magazines relate to Canon EOS cameras. We further assume that there are a
fixed number of people that will \textit{only} click on Group 1 adds (we
call them \textquotedblleft Group 1 Users"), and they click on each ad $i$
in Group 1 with a probability proportional to $q_{i}s_{i}$. Assume there are
a fixed number of people that might click on \textit{any of the ads }shown,
with a probability proportional to each ad's $q_{i}s_{i}$ . Based on this
click-through rate model, the swap of the positions of any two advertisers
will result in the change of click-through rate of all other advertisers and
our algorithm can no longer guarantee a social optimal ranking every time. }

{\normalsize We set the number of Group 1 and Group 2 Advertisers to be $6$
and $3$, respectively; set the number of Group 1 and Group 2 Users to be $%
400 $ and $200$, respectively. We ran the algorithm on $100$ auctions based
on the modified click-through rate model with randomly generated parameters $%
(v,~q,~e)$. We have obtained a $91.6\%$ averaged ratio of $\Pi _{se}(\bar{%
\sigma})$ and $\Pi _{se}(\sigma ^{\ast })$ for all runs, with a standard
deviation of $0.0549$. Although it is less than the averaged ratio we get
from the product click-through rate model (almost always larger than $99\%$%
), it can be seen as a reasonably satisfactory result. }

{\normalsize \bigskip }

\subsection{\protect\normalsize Efficiency of the Algorithm}

The proof of Theorem 3 shows that the optimal rank searching module in the
proposed algorithm takes {\normalsize $O\left( N^{2}\right) $ total number
of ad-quality score adjustments before the algorithm terminates. While the
number of score adjustments in the value\ revealing module depends on the
initial scoring and increment size of the score adjustment, it is polynomial
in terms of the input size }$v_{i}q_{i}$, for $i=1,...,N$. {\normalsize \ \
The biggest uncertainty that affects the efficiency of the algorithm comes
from the function }$NE\left( \cdot \right) ${\normalsize . In practice, in
order for the revenue to stabilize to determine whether a swap of the
neighboring advertisers is profitable, it is required that every time after
the scoring profile is adjusted, the advertisers form a certain bidding
equilibrium. We argue that over time, the advertisers are likely to learn
all relevant information to make an equilibrium decision. Furthermore, this
equilibrium does not necessarily need to be an NE and therefore does not
require each advertiser to have complete information about others' behavior.
In our implementation with complete information assumption, the algorithm
finds the NEs very quickly.}

\section{\protect\normalsize Conclusions and discussions}

{\normalsize 1. The proposed algorithm does not involve any functionality
that can not be implemented by the search engines. }

{\normalsize 2. The proposed algorithm does not involve any estimation of
the click-through rates but is solely driven by the total revenue generated
which can be observed directly by the search engine. In order for the
revenue to stabilize, it is required that every time after the scoring
profile is adjusted the advertisers form a certain bidding equilibrium. This
equilibrium does not necessarily need to be an NE and therefore does not
require each advertiser to have complete information about others' behavior.
}

{\normalsize 3. Both our theoretical and empirical analysis shows that the
search engine can extract almost all the surpluses from the advertisers,
given that there are more advertisers than slots and the values of the
advertisers can be accurately estimated. Though in reality some of the
arguments and assumptions in the model may not hold, our analysis
demystifies what a revenue-maximizing scoring strategy for the search engine
should look like. Specifically, it first needs to get close to an equalizing
scoring profile which solely depends on the estimation of advertisers'
values. Then by making infinitesimal differences between advertisers'
ranking scores, the ranking should be made in the order of the product of
the ad-quality specific factor and the advertiser's value. This is fully
consistent with what search engines interpret the ad-quality score. For
example, according Google AdWords, the \textquotedblleft Quality Score is
based on: 1) }The historical CTR of the ad on this and similar sites; 2) The
quality of your landing page".{\normalsize \ }

{\normalsize 4. The theoretical and empirical analysis shown in this paper
implies that in a monopoly market where advertisers do not have a choice of
other search engines, the monopoly search engine can abuse its power by
maximally extracting the surpluses from the advertisers. There are concerns
that \textquotedblleft ... since the ad quality factor is under the search
engine's control, it gives the search engine nearly unlimited power to
affect the actual ordering of the advertisers for a given set of bids" \cite%
{Easley10}. However, to our knowledge, no previous literature has studied
theoretically and empirically how far this \textquotedblleft unlimited
power" can go for a monopoly search engine. The analysis in this paper may
lead to new ways of identifying the abusage of a search engine's monopoly
power in charging the advertisers. For example, we can check whether the
\textquotedblleft ad-quality scores" do reflect the relative qualities in
the ads as they claimed to be or they are merely manipulating tools used by
the search engine to extract as much surpluses as possible from the
advertisers.\ }

%
%
%
%

{\normalsize
\bibliographystyle{IEEEbib}
\bibliography{ref}

\begin{thebibliography}{1}

\bibitem{Var09}
H.~R. Varian,
\newblock ``Online ad auctions,''
\newblock {\em Manuscript in preparation, UC Berkerly and Google}, 2009.

\bibitem{Edelman07}
B.~Edelman, M.~Ostrovsky, and M.~Schwarz,
\newblock ``Internet advertising and the generalized second-price auction:
  Selling billions of dollars worth of keywords,''
\newblock {\em American Economic Review}, vol. 97, no. 1, pp. 242--259.

\bibitem{Var07}
H.~R. Varian,
\newblock ``Position auctions,''
\newblock {\em International Journal of Industrial Organization}, vol. 25, pp.
  1163--1178, 2007.

\bibitem{Lahaie06}
S.~Lahaieh,
\newblock ``An analysis of alternative slot auction designs for sponsored
  search,''
\newblock in {\em Proceedings of the 7th ACM conference on Electronic
  Commerce}.

\bibitem{Easley10}
D.~Easley and J.~Kleinberg,
\newblock ``Networks, crowds, and markets: Reasoning about a highly connected
  world,''
\newblock {\em Cambridge University Press.}, 2010.

\end{thebibliography}
}

\end{document}